\documentclass[12pt]{article} 
\pdfoutput=1
\usepackage{amssymb,graphicx}
\usepackage{epstopdf}
\usepackage{amsmath,amsfonts}
\usepackage{epsfig} 
\usepackage{graphicx,graphics}
\usepackage{color}

\begin{document} 

\title{\bf Spectrum of gravitational waves from\\ long-lasting primordial sources}
\author{Sabir Ramazanov\\
\small{\em CEICO, Institute of Physics of the Czech Academy of Sciences,}\\
\small{\em Na Slovance 1999/2, 182 00 Prague 8, Czech Republic}
}

 \newcommand{\sabir}[1]{\textcolor{blue}{ #1}}

{\let\newpage\relax\maketitle}

\begin{abstract}
We discuss long-lasting gravitational wave sources arising and operating during radiation-dominated stage. Under a set of assumptions, we establish the correspondence between 
cosmological evolution of a source and the resulting gravitational wave spectrum. Namely, for the source energy density $\rho_s$ falling as a power law characterized by the exponent $\beta$, i.e., 
$\rho_s \propto 1/a^{\beta}$, where $a$ is the Universe scale factor, 
the spectrum takes the form $\Omega_{gw} \propto f^{2\beta-8}$ in certain ranges of values of constant $\beta$ and frequencies $f$. 
In particular, matching to the best fit power law shape of stochastic gravitational wave background discovered recently by Pulsar Timing Array collaborations, one identifies $\beta \approx 5$. 
We demonstrate the correspondence with concrete examples of long-lasting sources: domain walls and cosmic strings.

\end{abstract}

\sloppy

\section{Introduction}

 Recently various Pulsar Timing Array (PTA) collaborations including NANOGrav, PPTA, EPTA with InPTA, and CPTA, reported a signal consistent with stochastic gravitational wave (GW) background in the nHz frequency range~\cite{NANOGrav:2023gor, Reardon:2023gzh, EPTA:2023fyk, Xu:2023wog, InternationalPulsarTimingArray:2023mzf}. 
The spectral index $p$ of the best-fitting power law to the measured frequency $f$ dependence of the relic GW abundance, 
\begin{equation}
\Omega_{gw} \propto f^{p} \; ,
\end{equation}
is given by $p=1.8 \pm 0.6$ at $68\%$~CL~\cite{NANOGrav:2023hvm}\footnote{This corresponds to $\gamma=3.2 \pm 0.6$ in terms of the spectral index $\gamma \equiv 5-p$ commonly used by PTA collaborations.}. The latter is in tension with the value characteristic for GW-driven supermassive black hole binaries, $p=2/3$~\cite{Phinney:2001di}, or the spectral index describing infrared tails of some typical primordial sources, $p=3$ (see below). 
One should note, however, that current observational and modelling uncertainties are large enough, so that ``canonical'' GW sources like supermassive black hole binaries~\cite{Kocsis:2010xa, Sesana:2008mz, NANOGrav:2023hfp, Ellis:2023dgf} or first order phase transitions~\cite{NANOGrav:2023hvm, EPTA:2023xxk, Caprini:2019egz} still have a great potential of explaining the signal. (See also Ref.~\cite{Ellis:2023oxs} discussing various sources of primordial GWs in light of PTA data and the reference list there.) 
In particular, NANOGrav and EPTA collaborations~\cite{NANOGrav:2023hfp, EPTA:2023xxk} suggest a whole range of spectral indices for a background from supermassive black hole binaries. Hopefully, with the increased sensitivity of PTA data as well as addition of the new data, e.g., from LISA, one will be able to pinpoint characteristics of the stochastic GW background. Meanwhile, 
it is worth exploring, how the information about GW characteristics, i.e., the spectral index, can be used to constrain the properties of corresponding GW sources. This is the goal, which we partially 
address in the present work.

We focus on cosmological GW sources operating at radiation domination and 
fully relying on post-inflationary physics\footnote{Namely, we do not discuss GWs produced by second order scalar fluctuations enhanced at the end of inflationary stage~\cite{Matarrese:1997ay}. 
In that case one can obtain $p \approx 2$, provided that the scalar power spectrum exhibits a 
strong peak-like feature~\cite{Saito:2009jt}.}. In this situation, $p=3$ is perhaps the most common outcome supported by causality arguments, 
with an important loophole that it assumes a very short (in terms of a Hubble time) duration of  a source~\cite{Caprini:2009fx, Cai:2019cdl, Hook:2020phx}.

This suggests a way to obtain the spectral index $p \neq 3$ by switching to enduring GW sources lasting for many Hubble times. In particular, the value $p \approx 2$ favoured by the PTA data describes GW emission from the network of melting domain walls characterized by a time-dependent tension~\cite{Vilenkin:1981zs, Babichev:2021uvl,  Ramazanov:2021eya, Babichev:2023pbf}. 
While this GW source relies on a concrete particle physics scenario, only a few of its properties are relevant for obtaining the spectral index $p=2$, with the most non-trivial 
one being cosmological evolution of its energy density $\rho_s  \propto 1/a^5(t)$.
In the present work, we establish this on general grounds without resorting to melting domain walls and details of microscopic physics. More broadly, we establish the correspondence between the spectral index $p$ and the time-dependence $\rho_s=\rho_s (t)$ 
chosen to be of the form
\begin{equation}
\label{base}
\rho_s \propto \frac{1}{a^{\beta} (t)} \; ,
\end{equation}
where the exponent $\beta$ defining the redshift of the energy density $\rho_s$ is assumed to be constant.
We show that in a certain range of values of the constant $\beta$ and a range of frequencies (which is infinitely large for an everlasting GW source), one has 
\begin{equation}
\label{pbeta}
p=2\beta -8 \; ,
\end{equation} 
under a set of reasonable assumptions. The correspondence~\eqref{base} typically gets violated starting from\footnote{Generically, there is also a lower bound on $\beta$, for which the correspondence~\eqref{pbeta} holds, as it will 
be discussed in what follows.} $\beta=11/2$, when the standard spectral index $p=3$ is recovered: 
for $\beta>11/2$ the GW source effectively acts as instantaneous, and the value $p=3$ persists. 


\section{Generalities}

 Focusing on spin-2 linear perturbations of the FLRW metric, one writes
\begin{equation}
ds^2=a^2 (\tau) \left(-d\tau^2+[\delta_{ij}+h_{ij}]dx^{i} dx^{j} \right) \; ,
\end{equation}
where $h_{ij}$ is a transverse traceless tensor. Hereafter, we switch to conformal time $\tau$. The equation governing production of GWs in the presence of a source is given by 
\begin{equation}
\label{basic}
\left(\frac{\partial^2}{\partial \tau^2} +2aH\frac{\partial}{\partial \tau} -\frac{\partial^2}{\partial {\bf x}^2} \right) h_{ij} =16\pi G a^2 \rho_s (\tau) \Pi_{ij} ({\bf x}, \tau)  \; ,
\end{equation}
where $G$ is the Newton's constant. As in Ref.~\cite{Caprini:2009fx}, we performed separation of the source into the homogeneous background part $\rho_s (\tau)$ and dimensionless inhomogeneous part $\Pi_{ij} ({\bf x, \tau})$, which is also a transverse traceless tensor. In the next section, we will discuss how this separation can be performed in a non-ambiguous way, and at the moment treat it as a formal procedure. Let us switch to Fourier transform fields: 
\begin{equation}
\label{Fourier}
h_{ij} ({\bf x}, \tau)=\int d{\bf k} e^{i{\bf kx}} h_{ij} ({\bf k}, \tau) \qquad \Pi_{ij} ({\bf x}, \tau)=\int d{\bf k} e^{i{\bf kx}} \Pi_{ij} ({\bf k}, \tau) \; ,
\end{equation}
and perform splitting over polarizations: 
\begin{equation}
h_{ij} ({\bf k}, \tau)=\sum_{A} e^{A}_{ij} ({\bf k}) h_A ({\bf k}, \tau) \qquad  \Pi_{ij} ({\bf k}, \tau)=\sum_{A} e^{A}_{ij} ({\bf k}) \Pi_A ({\bf k}, \tau)\; .
\end{equation}
Here $e^{A}_{ij}$ is a pair of polarization tensors normalized as $e^{A}_{ij} e^{A'}_{ij}=2\delta_{AA'}$, where $A, A'=1,2$. We assume that no GW production takes place before $\tau_i$, when the source is switched on. 
By continuity of the field $h({\bf k}, \tau)$ and its time derivative $\partial h({\bf k, \tau})/ \partial \tau$ at $\tau_i$, one should impose initial conditions $h({\bf k}, \tau_i)=0$ and $\partial h({\bf k}, \tau)/\partial \tau |_{\tau=\tau_i}=0$. The solution of Eq.~\eqref{basic} in the background dominated by radiation reads
\begin{equation}
\label{solution}
h_A ({\bf k}, \tau)=\frac{16\pi G}{a(\tau) k} \cdot \int^{\tau}_{\tau_i} \sin k(\tau-\tau') \rho_s (\tau') a^3(\tau') \Pi_A ({\bf k}, \tau') d\tau' \; .
\end{equation}
This expression also holds after matter-radiation equality $\tau_{eq}$ in the most interesting case, when the modes enter the horizon at radiation-domination, so that $k\tau_{eq} \gg 1$. The energy density of GWs today is given by 
\begin{equation}
\label{densitytoday}
\rho_{gw} ({\bf x}, \tau_0) =\frac{1}{32\pi G a^2_0} \left \langle \frac{\partial h_{ij} ({\bf x}, \tau)}{\partial \tau} \frac{\partial h_{ij} ({\bf x}, \tau)}{\partial \tau} \right \rangle_0 \; ,
\end{equation}
where the subscript $'0'$ refers to the present Universe, and by $\langle ... \rangle$ we mean the ensemble average over many realizations of the stochastic source. Assuming that the source is terminated at some time $\tau_f \ll \tau_0$, using Eqs.~\eqref{Fourier} and~\eqref{solution}, we obtain for the GW energy density:
\begin{equation}
\begin{split}
\label{long}
\rho_{gw} ({\bf x}, \tau_0) &=\frac{4\pi G}{a^4_0} \sum_{A,A'}\int d{\bf k} d{\bf p} \; e^{A}_{ij} ({\bf k})e^{A'}_{ij} ({\bf p}) \; e^{i{\bf kx}+i{\bf p}{\bf x}}  \int^{\tau_f}_{\tau_i} d\tau' \int^{\tau_f}_{\tau_i} d\tau'' \rho_s (\tau') \rho_s (\tau'') a^3 (\tau') a^3(\tau'') \times \\ 
& \times \langle \Pi_{A} ({\bf k}, \tau') \Pi_{A'} ({\bf p}, \tau'') \rangle \cdot [\cos \left(p\tau_0 -k\tau_0 +k\tau'-p\tau'' \right)  +\cos \left(
p\tau_0 +k\tau_0 -k\tau'-p\tau'' \right) ]\; .
\end{split}
\end{equation}
Note that we limit both $\tau_i$ and $\tau_f$ to be within radiation era, with no much loss of generality. 
Indeed, PTAs as well as other current or planned GW observatories operate in a range of frequencies characteristic for GW emission taking place at radiation domination for not very small reheating temperatures.  
 Assuming statistically homogeneous, isotropic, and unpolarized sources, so that
\begin{equation}
\label{correlator}
\langle \Pi_A ({\bf k}, \tau) \Pi_{A'} ({\bf p}, \tau')\rangle =(2\pi)^3 \delta_{AA'} \delta ({\bf k}+{\bf p}) P (k, \tau, \tau') \; ,
\end{equation}
omitting\footnote{One can estimate that the second cosine in Eq.~\eqref{long} experiences $\sim 6\cdot 10^{9} \cdot (\delta k/(2\pi a_0~\mbox{nHz}))$ oscillations in a frequency range $\delta k/(2\pi a_0)$. This is a huge number for 
any experimentally accessible range $\sim \delta k/(2\pi a_0)$. In particular, resolving these oscillations with PTAs or gravitational interferometers is impossible, which justifies omitting the fast oscillating cosine in Eq.~\eqref{interm}.} the fast oscillating cosine function in Eq.~\eqref{long}, and summing over polarizations of GWs, we obtain the spectral energy density:
\begin{equation}
\label{interm}
\frac{d\rho_{gw}}{d\ln k}=\frac{512\pi^5 G k^3}{ a^4_0} \cdot  \int^{\tau_f}_{\tau_i} d\tau' \int^{\tau_f}_{\tau_i} d\tau'' \rho_s (\tau') \rho_s (\tau'') a^3 (\tau') a^3 (\tau'') \cos \left(k[\tau'-\tau''] \right)P (k, \tau', \tau'') \; .
\end{equation}
Here we introduced the power spectrum of the source $P(k, \tau', \tau'')$, which will play a key role in what follows. 

\section{Spectral index of gravitational waves}
 The form of the power spectrum $P (k, \tau', \tau'')$ can be fixed from its mass dimension and scaling properties. Namely, let us perform a transformation upon the scale factor $a \rightarrow \zeta a$ supplemented 
by coordinate transformations ${\bf k} \rightarrow \zeta {\bf k}$ and $\tau \rightarrow \tau/\zeta$ with $\zeta>0$ being some constant, so that the physical quantities $k/a$ and $a \tau$ remain invariant. It is straightforward to check that the function $P(k, \tau, \tau')$ transforms under the rescaling as 
\begin{equation}
\label{rescalingP} 
P(k, \tau, \tau') \rightarrow \frac{P(k, \tau, \tau')}{\zeta^3} \; ,
\end{equation}
which is traced back to the fact that the dimensionless stress $\Pi_{ij} ({\bf x}, \tau)$ is invariant under the rescaling. This, in turn, can be inferred from Eq.~\eqref{basic}. 
Keeping in mind the mass dimension of the power spectrum $P(k, \tau, \tau')$, which is minus three, one can write:
\begin{equation}
\label{scaling}
P(k, \tau', \tau'')=\frac{{\cal P} (k\tau', k\tau'')}{k^3}\; ,
\end{equation}
where ${\cal P}$ is the reduced dimensionless power spectrum depending only on combinations of coordinates $k\tau'$ and $k\tau''$. Using Eqs.~\eqref{Fourier},~\eqref{correlator}, and ~\eqref{scaling}, one can show 
that the contribution of superhorizon modes obeying $k\tau \leq 1$ to the tensor $\Pi_{ij} ({\bf x}, \tau)$ is described by
\begin{equation}
\label{separation}
\langle \Pi^2_{ij} ({\bf x}, \tau) \rangle \left. \right|_{k\tau \leq 1} =128\pi^4  \int^{1}_0 \frac{d (k\tau)}{k\tau} {\cal P} (k\tau, k\tau) \; .
\end{equation}
We observe that the quantity $\langle \Pi^2_{ij} ({\bf x}, \tau) \rangle \left. \right|_{k\tau \leq 1}$ is constant, provided that the integral on the r.h.s. here is convergent, which will be proven later in this section. 
This means that cosmological evolution of the GW source term on the r.h.s. of Eq.~\eqref{basic}, i.e., $\rho_s (\tau) \Pi_{ij} ({\bf x}, \tau)$, is indeed determined by $\rho_s (\tau)$. 
Hence, the choice of the power spectrum~\eqref{scaling} eliminates ambiguities regarding the separation into the background $\rho_s (\tau)$ and tensorial parts $\Pi_{ij} ({\bf x}, \tau)$.

However, the expression~\eqref{scaling} is not the most generic one, because the function ${\cal P}$ may depend on the initial time of GW emission $\tau_i$ as well as mass parameters $M_n (\tau)$ (which include the Hubble rate and generically can depend on time) characteristic for a concrete model\footnote{Note that in Eq.~\eqref{generic} we do not write the dependence on the ratio $k/(a M_n)$. The reason is that one can rewrite $k/(a H) \cdot (H/M_n) = k \tau \cdot (H/M_n)$, which is already included in Eq.~\eqref{generic}.}:
\begin{equation}
\label{generic}
{\cal P}={\cal P} \left(k\tau', k\tau'', k\tau_i, \{ M_l/M_n \}\right) \; ,
\end{equation}
where $\{ M_l/M_n \}$ with $l \neq n$ denotes the set of all possible ratios of masses $M_n$. To obtain Eq.~\eqref{scaling}, we first eliminate dependence on $\tau_i$ by simply stating that
\begin{equation}
\label{splitting}
\frac{\partial {\cal P}}{\partial \tau_i}=0 \; .
\end{equation}
This is a non-trivial assumption, and in the case of topological defects 
it is a manifestation of scaling regime. The scaling regime is indeed observed for cosmic strings~\cite{Vanchurin:2005pa, Martins:2005es, Ringeval:2005kr} and domain walls~\cite{Press:1989yh, Hiramatsu:2013qaa}, but one should be careful when applying 
Eq.~\eqref{splitting} to generic sources. Regarding the masses $M_n$, it is enough to consider time-varying ratios $M_l/M_n$, because for constant $M_l/M_n$ the form of Eq.~\eqref{scaling} is clearly not affected. Dependence on $M_l /M_n $ is innocuous 
if it can be factored out, i.e., provided that ${\cal P}={\cal P}_1 (M_l/M_n) \cdot {\cal P}_2 (k\tau', k\tau'')$. In this case, we can trivially redefine ${\cal P} \equiv {\cal P}_2$ and at the same time absorb the dependence on $M_l/M_n$ into the background energy density 
$\rho_s (\tau)$. Otherwise, we should make an additional assumption to ensure that the ratios $M_l/M_n$ do not spoil Eq.~\eqref{scaling}. Naturally one can assume that different masses $M_l (\tau)$ and $M_n (\tau)$ are hierarchically ordered, i.e., 
$M_l \ll M_n$ with no loss of generality, or become so after some time. We require that the power spectrum ${\cal P}$ remains a well-behaved function taking finite values in the limit $M_l/M_n \rightarrow 0$:
\begin{equation}
\label{assumption2}
0<{\cal P} \left(k\tau', k\tau'', \{M_l/M_n \} \right)  \left. \right|_{M_l/M_n \rightarrow 0} <\infty \; .
\end{equation}
For example, in the case of domain walls, there are two mass parameters describing the system: the Hubble rate $H \equiv M_1$ and inverse wall width $\delta^{-1}_w \equiv M_2$. 
As the ratio $M_1/M_2$ decreases (walls become thin relative to the Hubble radius), the system enters the scaling regime characterized by the unique parameter $H$~\cite{Hiramatsu:2013qaa}, 
so that the source power spectrum takes the form~\eqref{scaling}.

Assuming Eqs.~\eqref{splitting} and~\eqref{assumption2}, we proceed with Eq.~\eqref{scaling}. Substituting Eqs.~\eqref{base} and~\eqref{scaling} into Eq.~\eqref{interm} and defining 
\begin{equation}
\xi \equiv k\tau \; , 
\end{equation}
we obtain
\begin{equation}
\label{almost}
\frac{d\rho_{gw}}{d\ln k}=\frac{512\pi^5 G k^{2\beta-8}\rho^2_s (\tau_i)a^{6} (\tau_i) }{a^4_0 \tau^{6-2\beta}_i} \cdot  \int^{\xi_f}_{\xi_i} \int^{\xi_f}_{\xi_i}d\xi' d\xi'' (\xi' \xi'')^{3-\beta} \cos (\xi'-\xi'') {\cal P} (\xi', \xi'') \; .
\end{equation}
Recall that the constant $\beta$ describes the redshift of the source energy density, i.e., $\rho_s (\tau) \propto 1/a^{\beta} (\tau)$. The expression in front of the double integral here is independent of $\tau_i$; this is clear from Eq.~\eqref{base} and the fact that $a(\tau) \propto \tau$ during radiation domination. Furthermore, this expression already yields 
the spectral index~\eqref{pbeta} quoted in the introduction. 
Nevertheless, additional dependence on the wavenumber $k$ as well as times $\tau_i$ and $\tau_f$ may enter through integration limits $\xi_i$ and $\xi_f$. One can ignore this dependence, provided that the integral is saturated at some $\xi'$ and $\xi''$ inside the region $(\xi_i, \xi_f)$, so that 
\begin{equation}
\label{assump1}
\int^{\xi_f}_{\xi_i} \int^{\xi_f}_{\xi_i}d\xi' d\xi'' (\xi' \xi'')^{3-\beta} \cos (\xi'-\xi'') {\cal P} (\xi', \xi'') \approx \int^{+\infty}_{0} \int^{+\infty}_{0}d\xi' d\xi'' (\xi' \xi'')^{3-\beta} \cos (\xi'-\xi'') {\cal P} (\xi', \xi'') \; .
\end{equation}
We treat Eq.~\eqref{assump1} as our new assumption for a while, and later on recast it into constraints on the constant $\beta$. In particular, Eq.~\eqref{assump1} implies that the integral on the r.h.s. of Eq.~\eqref{almost} is finite:
\begin{equation} 
\label{convergence}
 \int^{+\infty}_{0} \int^{+\infty}_{0}d\xi' d\xi'' (\xi' \xi'')^{3-\beta} \cos (\xi'-\xi'') {\cal P} (\xi', \xi'') \equiv C<\infty \; .
\end{equation}
The constant $C$ here is independent of $k$; this property can be traced back to our assumptions~\eqref{splitting} and~\eqref{assumption2}. Hence, in the range of momenta~\eqref{range}, we can write in terms of 
GW relic abundance, 
\begin{equation}
\label{final}
\Omega_{gw} (f)= \frac{2 (2\pi)^{2\beta} C G f^{2\beta-8} \rho^2_s (\tau_i) }{\pi^3 (a_0\tau_i)^{6-2\beta}} \cdot \left(\frac{a(\tau_i)}{a_0} \right)^6 \; ,
\end{equation}
which explains Eq.~\eqref{pbeta}. Here $\Omega_{gw}$ is defined as 
\begin{equation}
\Omega_{gw} (f) =\frac{1}{\rho_{tot,0}} \cdot \frac{ d\rho_{gw}}{ d \ln f} \; ,
\end{equation}
where $\rho_{tot, 0}$ is the energy density of the Universe today, and the frequency $f$ is related to the conformal momentum $k$ by 
\begin{equation}
f=\frac{k}{2\pi a_0} \; .
\end{equation}
For the particular value of the parameter $\beta =5$, we end up with
\begin{equation}
\label{best}
\Omega_{gw} \propto f^2 \; ,
\end{equation}
which is the behaviour favoured by the recent PTA data. Were the condition~\eqref{assump1} violated, the spectral energy density would depend on $\xi_i$ and $\xi_f$, which would yield the additional frequency $f$ dependence eventually spoiling Eqs.~\eqref{final} and~\eqref{best}. 

Validity of Eq.~\eqref{assump1} implies restriction on the range of momenta $k$, or equivalently frequencies $f$, which cannot be arbitrary for fixed $\tau_i$ and $\tau_f$. Generically, one writes
\begin{equation}
\label{range}
\frac{\alpha}{2\pi a_0\tau_f} \ll f \ll \frac{\alpha}{2\pi a_0\tau_i} \; , 
\end{equation}
where the coefficient $\alpha \gtrsim 1$ relates the characteristic frequency of GW emission at some time $\tau$ to the Hubble rate $H(\tau)$ (we give a more precise definition shortly). The coefficient $\alpha$ is constant for topological 
defects in the scaling regime~\cite{Hiramatsu:2013qaa, Blanco-Pillado:2013qja} meaning that the characteristic frequency 
is pinned to $H(\tau)$ in that case. Note, however, that our further discussion does not rely on this property, i.e., it is also applicable to the case of time-varying $\alpha$. For frequencies $f$ violating Eq.~\eqref{range}, the most efficient GW emission occurs outside of the region $(\xi_i, \xi_f)$, and Eq.~\eqref{assump1} cannot be fulfilled. The range~\eqref{range} defines desired duration of a GW source, which should be long enough to cover frequency bands of observational facilities, e.g., PTAs. For the particular value $\alpha \simeq 2\pi$, which is characteristic for domain walls~\cite{Hiramatsu:2013qaa}, the NANOGrav range $10^{-9}~\mbox{Hz} \lesssim f \lesssim 5 \times 10^{-8}~\mbox{Hz}$ corresponds to the GW source, 
which operates at least during the times, when the Universe cools down from the temperature $T \sim 1~\mbox{GeV}$ to $T \sim 20~\mbox{MeV}$.

In the remainder of the section, we find the conditions for which Eq.~\eqref{assump1} is justified. If GW emission at each moment $\tau$ occurred at some particular frequency $f$, the approximation~\eqref{assump1} would be valid for arbitrary $\beta$. Commonly, however, one deals 
with a spectrum of frequencies.  
Nevertheless, we can show that there is a rather broad range of values of the parameter $\beta$, where the conditions~\eqref{assump1} and~\eqref{convergence} hold, at least for two well-motivated choices of the function ${\cal P} (\xi', \xi'')$. 
To find this range, let us focus on GWs emitted during the short time interval $\tau_1 \leq \tau \leq \tau_2$, so that $k(\tau_2-\tau_1) \ll 1$ for all relevant momenta. We approximate this transient GW emission by the broken power law described by the peak frequency $f_p  \simeq \alpha/(2\pi a_0 \tau)$ (this expression defines the coefficient $\alpha$ 
entering Eq.~\eqref{range}) separating infrared (IR) and ultraviolet (UV) tails:
\begin{align}
\label{instspectrum}
\frac{d\rho_{gw}}{d\ln f} \propto 
\begin{cases}
f^3 \qquad 2\pi a_0 f \tau \ll \alpha\\
f^{q} \qquad 2\pi a_0 f \tau \gg \alpha \; ,
\end{cases}
\end{align}
where $q$ is a constant characterizing the UV tail. The latter is not robust against details of a particular GW source, and this is the reason, why we choose to do not specify $q$. 
For example, (constant tension) domain walls~\cite{Hiramatsu:2013qaa} or bubble collisions during first order phase transitions~\cite{Caprini:2019egz} lead to $q=-1$; at the same time 
acoustic waves triggered by phase transitions lead to $q=-3$~\cite{Hindmarsh:2016lnk, Hindmarsh:2017gnf}. 
 On the other hand, the IR tail $\propto f^3$ fixed by causality considerations is a rather generic property of transient sources. This universality is, however, violated in certain 
 cases, cf. Refs.~\cite{Cai:2019cdl, Hook:2020phx, Brzeminski:2022haa}, and we will encounter with one situation of that kind in what follows. Note that if the actual IR tail is steeper than in Eq.~\eqref{instspectrum}, 
 our discussion below is still applicable in a conservative sense. Having this said, we continue with Eq.~\eqref{instspectrum}.
 
 Let us first assume the case of a totally uncorrelated source~\cite{Caprini:2009fx}: 
 \begin{equation}
 \label{uncorr}
 {\cal P} (\xi', \xi'')= {\cal F}(\xi') \delta (\xi'-\xi'') \; .
 \end{equation}
 Substituting this into the double integral in Eq.~\eqref{almost} and restricting to the short time interval $(\tau_1, \tau_2)$ as described above, we obtain 
 \begin{equation}
\label{doubleuncorr}
\int^{\xi_{2}}_{\xi_1} \int^{\xi_2}_{\xi_1} d\xi' d\xi'' (\xi' \xi'')^{3-\beta} \cos (\xi'-\xi'') {\cal P} (\xi', \xi'') \sim {\cal F}(\xi_1) \cdot \left(\xi^{7-2\beta}_2 -\xi^{7-2\beta}_1 \right) \; . 
\end{equation}
Looking at Eq.~\eqref{almost}, we observe that recovering the IR tail $\propto f^3$ requires ${\cal F} (\xi_1) \propto \xi^4_1$ for $\xi_1 \ll \alpha$. (Recall that $\xi \equiv k\tau \equiv 2\pi a_0 f \tau$). 
Since there is no preferred choice of $\xi_1$, we can write 
\begin{equation}
{\cal F} (\xi') \propto \xi^{'4} \qquad    \xi' \ll \alpha   \; .
\end{equation}
Substituting this into Eq.~\eqref{almost}, one finds values of $\beta$, for which it is legitimate to take the limit $\xi_i \rightarrow 0$ in the double integral in Eq.~\eqref{almost} consistently with our assumption~\eqref{assump1}. 
This occurs for $\beta <\frac{11}{2}$. Apparently, the case of our major interest $\beta=5$ fulfils this condition. In a similar manner, one could infer the lower bound on the constant $\beta$ from the behaviour of the UV tail $\propto f^q$. 
Again restricting to the instantaneous source obeying $\xi_2-\xi_1 \ll 1$, but now assuming $\xi_1,~\xi_2 \gg \alpha$, we obtain
\begin{equation}
{\cal F} (\xi') \propto \xi'^{q+1} \qquad    \xi' \gg \alpha \; .
\end{equation}
Hence, convergence of the integral~\eqref{almost} in the limit $\xi_f \rightarrow \infty$ gives $\beta >3+q/2$. 

Now let us consider a totally correlated source corresponding to the following choice of the power spectrum~\cite{Caprini:2009fx}:
\begin{equation}
\label{corr} 
{\cal P} (\xi', \xi'') = \sqrt{ {\cal P} (\xi', \xi') \cdot  {\cal P} (\xi'', \xi'')} \; .
\end{equation}
Then, in the short time interval $(\tau_1, \tau_2)$, the double integral in Eq.~\eqref{almost} takes the form 
\begin{equation}
\label{double}
\int^{\xi_{2}}_{\xi_1} \int^{\xi_2}_{\xi_1} d\xi' d\xi'' (\xi' \xi'')^{3-\beta} \cos (\xi'-\xi'') {\cal P} (\xi', \xi'') \sim {\cal P} (\xi_1, \xi_1) \cdot \left(\xi^{4-\beta}_2 -\xi^{4-\beta}_1 \right)^2 \; . 
\end{equation}
 As it follows from Eq.~\eqref{almost}, recovering the IR tail $\propto f^3$ requires ${\cal P} (\xi_1, \xi_1) \propto \xi^3_1$ for $\xi_1 \ll \alpha$. In view of Eq.~\eqref{corr}, this gives 
\begin{equation}
\label{corrsepar}
{\cal P} (\xi', \xi'') \propto (\xi' \xi'')^{3/2} \qquad    \xi',~\xi'' \ll \alpha   \; .
\end{equation}
Let us note here that such behaviour of the power spectrum ${\cal P} (\xi', \xi'')$ for small $\xi'$ and $\xi''$ warrants convergence of the integral in Eq.~\eqref{separation}.
Consistency of Eq.~\eqref{corrsepar} with our assumption~\eqref{assump1} again yields the upper limit $\beta<11/2$. 
On the other hand, the lower bounds on $\beta$ are slightly different for totally uncorrelated and correlated sources. Indeed, in the latter case, one cannot disregard the cosine factor in the integrand of Eq.~\eqref{almost}, 
and consequently the constraint on the constant $\beta$ gets slightly relaxed: $\beta >3+q/2$.

 Conservatively, we summarize the range of values of the parameter $\beta$, for which the approximation~\eqref{assump1} is valid, 
as follows:
\begin{equation}
\label{constraints}
4+\frac{q}{2}<\beta <\frac{11}{2} \; .
\end{equation}
 In particular, for a not very blue UV tail of the transient source, $q< 2$, the correspondence between the value $\beta=5$ and the best fit PTA spectral index $p=2$ holds. Violations of Eq.~\eqref{constraints}, i.e., 
\begin{equation}
\label{away}
\beta \geq \frac{11}{2}~\mbox{or}~\beta \leq 4+\frac{q}{2} \; ,
\end{equation}
may suggest that GW emission is strongly saturated close to the initial moment $\tau_i$  or the endpoint $\tau_f$, respectively. In both cases, one typically approximates
the source as instantaneous, and results with the GW spectrum~\eqref{instspectrum}. Let us stress, however, that the first of Eq.~\eqref{away} 
is sensitive to the IR tail of the instantaneous GW emission (see the comment after Eq.~\eqref{instspectrum}), and the second of Eq.~\eqref{away} --- to the choice of the function ${\cal P} (\xi', \xi'')$.

Some important remarks are in order here. It is likely that the realistic behaviour of the source is interpolating between Eqs.~\eqref{uncorr} and~\eqref{corr}: in the regime $|\xi'-\xi''| \gg 1$, the approximation of an uncorrelated source~\eqref{uncorr} appears to be more suitable, while for $|\xi'-\xi''| \ll 1$ we expect that the source can be treated as sufficiently coherent, and one should rather use Eq.~\eqref{corr}. The constraints~\eqref{constraints} are valid also in this hybrid scenario, 
but the lower bound on $\beta$ can be slightly relaxed, i.e., $\beta >3+q/2$. This is again due to the cosine factor in the integrand of Eq.~\eqref{almost}, as it is described above. Keeping that in mind, we go ahead with our 
conservative constraints~\eqref{constraints}.

\section{Examples}
  Let us apply the correspondence $p=2\beta-8$ to some concrete examples of long-lasting sources. Perhaps the most well-known source of this type is represented by constant tension cosmic strings~\cite{Kibble:1976sj} resulting from spontaneous symmetry breaking of global or gauge continuous symmetries. In the scaling regime, the energy density of the network of cosmic strings during radiation domination mimics that of relativistic species, i.e., 
$\rho_s \propto 1/a^4 (\tau)$. Hence, $\beta=4$ and we end up with the flat spectrum $\Omega_{gw} (f) \propto f^{0}$~\cite{Vilenkin:1981bx, Vachaspati:1984gt, Vilenkin}. This conclusion can be extrapolated to a wide range of different topological defects with the same redshift behaviour as cosmic strings~\cite{Fenu:2009qf, Figueroa:2012kw, Figueroa:2020lvo}. 

On the other hand, the network of constant tension domain walls~\cite{Zeldovich:1974uw} resulting from spontaneous breaking of discrete symmetries has a different cosmological evolution $\rho_{s} \sim \sigma_{wall} H \propto 1/a^2 (\tau)$ in the scaling regime. Here $\sigma_{wall} =\mbox{const}$ is a wall tension (wall mass per unit area) related to the expectation value $v$ of the scalar field experiencing symmetry breaking by $\sigma_{wall} \propto v^3$. 
Consequently, one gets $\beta=2$, which does not obey~\eqref{constraints}, unless $|q|$ is chosen to be unnaturally large. As it follows from the comment after Eq.~\eqref{away}, this source effectively acts as an instantaneous one, and the GW spectrum 
is given by Eq.~\eqref{instspectrum} with the value $q=-1$ fixed by numerical simulations~\cite{Hiramatsu:2013qaa}. It is worth stressing that despite the spectral shape $\propto f^3$ is disfavoured by the PTA data, the broken power law~\eqref{instspectrum} with a sufficiently smooth behaviour around the peak frequency fits well observations~\cite{NANOGrav:2023hvm} given present error bars and modelling uncertainties. This motivates building particle physics scenarios, which involve 
domain walls, cf. Refs.~\cite{Madge:2023cak}.

Topological defects can also be realized in scale-invariant models~\cite{Babichev:2021uvl, Ramazanov:2021eya, Emond:2021vts}, in which case their tension decreases with cosmic expansion rather than set by constant mass parameters, as in the examples above. In particular, 
melting domain walls mentioned in the introduction are characterized by the tension $\sigma_{wall} \propto T^3(\tau)$~\cite{Vilenkin:1981zs, Babichev:2021uvl, Ramazanov:2021eya}. In this case, the energy density of the domain wall network 
behaves as $\rho_s \sim \sigma_{wall} H \propto 1/a^5(\tau)$ in the scaling regime, so that $\beta=5$, and Eq.~\eqref{pbeta} gives $p=2$, which is in agreement with the PTA data~\cite{Babichev:2023pbf}. Notably 
also, the energy density $\rho_s (\tau)$ decreases fast enough relative to radiation, and the infamous domain wall problem is avoided in this scenario.

Finally, we would like to mention the case of melting cosmic strings~\cite{Emond:2021vts} 
characterized by the time-dependent tension (string mass per unit length) $\mu \propto T^2(\tau)$. The energy density of the melting string network evolves with time according to $\rho_s \sim \mu H^2 \propto 1/a^{6} (\tau)$ corresponding 
to $\beta=6$. The latter does not fulfil the condition~\eqref{constraints}; the caveat, however, is that transient GW emission does not have the IR tail in the case of cosmic strings, at least for the choice of loop production 
function assumed in Ref.~\cite{Emond:2021vts}. That is, Eq.~\eqref{pbeta} is still applicable and one has $\Omega_{gw} \propto f^4$. This reiterates the statement that the constraints~\eqref{constraints} should be regarded as a conservative ones, 
and their violation does not necessarily invalidate the correspondence~\eqref{pbeta}.

\section{Conclusions}
 We have shown that the GW spectrum created by the enduring source with the energy density evolving as $\rho_s \propto 1/a^{\beta}$ has a power law behaviour described by the spectral index $p=2\beta-8$ in the frequency range~\eqref{range}, provided that the conditions~\eqref{scaling} and~\eqref{assump1} are fulfilled. The condition~\eqref{assump1} can be reinterpreted as constraints~\eqref{constraints} on the constant $\beta$, 
 at least for the well-motivated choices of the source power spectrum given by Eqs.~\eqref{uncorr} and~\eqref{corr}. Using the correspondence $p=2\beta-8$, one can extract information, which 
 is essential for identifying the source of GW background. It is worth mentioning, however, that the full problem of the source identification requires much more information, i.e., the knowledge of the 
 GW amplitude, full spectral shape, location of characteristic frequencies etc. However, understanding these GW properties is impossible within our model-independent analysis, and therefore it lies out of the scope of the present work.

Soft breaking of the condition~\eqref{scaling} is expected to induce small corrections into Eq.~\eqref{pbeta}. There are also inevitable departures from Eq.~\eqref{pbeta} due to dependence on 
the wavenumber $k$ of the integration limits in Eq.~\eqref{almost}, or equivalently due to the approximate nature of Eq.~\eqref{assump1}. While such departures are small within assumptions made, they can be important in view 
of future precision measurements of stochastic GW background. 
Furthermore, throughout the paper we neglected the change of relativistic degrees of freedom $g_* (T)$ during GW emission. This change of $g_* (T)$ is also expected to impact the spectral shape~\eqref{final}, 
which can be particularly prominent at the times of QCD phase transition~\cite{Franciolini:2023wjm}. 

The assumption~\eqref{splitting} underlying the functional form~\eqref{scaling} is perhaps 
the most restrictive one. In the future, it would be interesting to find situations, where it holds, besides topological defects in the scaling regime. 
In particular, long lasting effects triggered by first order phase transitions deserve a special attention. These effects are known to yield the GW spectrum $\Omega_{gw} \propto f$ in a particular range of frequencies~\cite{Hindmarsh:2016lnk, Hindmarsh:2017gnf, Jinno:2017fby, Konstandin:2017sat, Lewicki:2022pdb}. 
This may suggest time evolution of the corresponding source $\rho_s \propto 1/a^{9/2} (\tau)$, if the assumptions~\eqref{splitting} and~\eqref{assumption2} are approximately valid in this case.

In concluding, we wish to note that the present discussion motivated mainly by recent findings of PTA collaborations, may be also useful in view of forthcoming GW searches with LISA~\cite{LISA:2017pwj} and Einstein Telescope~\cite{Hild:2010id}. 
In this regard, it would be interesting to generalize results of this work to an arbitrary equation of state of the Universe $w$. While we focused on radiation domination 
with $w=1/3$, LISA and Einstein Telescope can be sensitive to earlier evolution stages for relatively low reheating temperatures.

{\it Acknowledgments.} We are indebted to Rome Samanta and Alexander Vikman for useful discussions. SR acknowledges the support of the Ministry of Education, Youth and Sports (MEYS) through the INTER-EXCELLENCE II, INTER-COST grant LUC23115.

\end{document}